\documentstyle[amssymb,aps,twocolumn,psfig]{revtex}

\begin{document}
\draft
\title{The microcanonical theory and pseudoextensive systems}
\author{L. Velazquez\thanks{%
luisberis@geo.upr.edu.cu}}
\address{Departamento de F\'{i}sica, Universidad de Pinar del R\'{i}o\\
Mart\'{i} 270, esq. 27 de Noviembre, Pinar del R\'{i}o, Cuba. }
\author{F. Guzm\'{a}n\thanks{%
guzman@info.isctn.edu.cu}}
\address{Departamento de F\'{i}sica Nuclear\\
Instituto Superior de Ciencias y Tecnolog\'{i}as Nucleares\\
Quinta de los Molinos. Ave Carlos III y Luaces, Plaza\\
Ciudad de La Habana, Cuba.}
\date{\today}
\maketitle

\begin{abstract}
In the present paper are considered the self-similarity scaling postulates
in order to extend the Thermodynamics to the study of one special class of
nonextensive systems: the pseudoextensive, those with exponential behavior
for the asymptotical states density of the microcanonical ensemble. It is
shown that this kind of systems could be described with the usual
Boltzmann-Gibbs' Distribution with an appropriate selection of the
representation of the movement integrals. It is shown that the
pseudoextensive systems are the natural frame for the application of the
microcanonical thermostatistics theory of D. H. E. Gross.
\end{abstract}


\section{Introduction}

In our previous work we addressed the problem of generalizing the extensive
postulates in order to extend the application of the Thermostatistics to
study of the nonextensive systems \cite{vel1}. Our proposition was that this
derivation could be carried out taking into consideration the
self-similarity scaling properties of the systems with the increasing of
their degrees of freedom and analyzing the conditions for the equivalence of
the microcanonical ensemble with some generalization of the canonical
ensemble in the thermodynamic limit (ThL). The last argument has a most
general character than the Gibbs' one, since it is not sustained by the
axiom of the separability of one subsystem from a whole system, invalid
supposition for the nonextensive systems due to the long-range correlations
that exist in them.

In order to apply the above considerations it must be taken into account the
geometrical aspects of the probabilistic distribution functions (PDF) of the
ensembles. In the ref.\cite{vel1} we showed that the microcanonical ensemble
possesses reparametrization invariance, $Diff\left( \Im _{N}\right) $, for
the movement integrals abstract space of the distribution, $\Im _{N}$. This
property is completely in contradiction with the ergodic chauvinism that
characterizes the traditional Thermodynamics. Under the supposition that the
generalized canonical ensemble could be derived from a Probabilistic
Thermodynamic Formalism (PThF), those based on the probabilistic
interpretation of the entropy concept, it is deduced that its PDF only
admits a certain set of representations of $\Im _{N}$ in agreement with the
self-similarity scaling properties of the system, ${\cal M}_{c}{\cal =}%
\left\{ \Re _{I}\right\} $, which are related by the transformations of the
Special Lineal Group, $SL\left( {\bf R}^{n}\right) $, where $n=\dim \left(
\Im _{N}\right) $. This fact suggests that a spontaneous symmetry breaking
takes place from $Diff\left( \Im _{N}\right) $ to $SL\left( {\bf R}%
^{n}\right) $ due to the scaling properties of the systems during the ThL,
that is, the ThL could be identified with the establishment of the
self-similarity scaling properties in the system.

The present paper will be devoted to the application of the above ideas to
the analysis of those systems possessing a simplest scaling laws: {\em the
exponential}. In this particular case, the self-similarity scaling laws of
the systems in the ThL are defined by:

\begin{equation}
\left. 
\begin{array}{c}
N\rightarrow N\left( \alpha \right) =\alpha N \\ 
I\rightarrow I\left( \alpha \right) =\alpha ^{\varkappa }I \\ 
a\rightarrow a\left( \alpha \right) =\alpha ^{\pi _{a}}a
\end{array}
\right\} \Rightarrow W_{asym}\left( \alpha \right) =\exp \left[ \alpha
^{\kappa }\ln W_{asym}\left( 1\right) \right] \text{,}
\end{equation}
where $W_{asym}$ is the accessible volume of the macroscopic state in the
system configurational space in the ThL, $\alpha $ is the scaling parameter, 
$\varkappa $ , $\pi _{a}$ and $\kappa $ are real constants characterizing
the scaling transformations, where: 
\begin{equation}
\text{\ }0<\kappa \leqslant 1\text{.}  \label{c1}
\end{equation}
Finally, $W_{asym}\left( \alpha \right) $ is given by:

\begin{equation}
W_{asym}\left( \alpha \right) =W_{asym}\left[ I\left( \alpha \right)
,N\left( \alpha \right) ,a\left( \alpha \right) \right] \text{.}
\end{equation}

The equality in the Eq.(\ref{c1}) can only take place when exist
non-correlated states in the ThL. It will be referred as {\it pseudoextensive%
} those systems with scaling laws given by the above conditions. The reason
for such denomination is easy to understand since the extensive properties
of the systems described by the traditional Thermodynamic are one special
case of the exponential scaling laws. It is easy to see that a great variety
of systems could be grouped under this denomination. In general, all the
Hamiltonian systems of identical particles with an additive kinetic part are
pseudoextensive independently of the interactions among them.

According to ref.\cite{vel1}, in this case it must be assumed the classical
expression of the {\bf Boltzmann's Principle }\cite{bol1,bol2,ein}:

\begin{equation}
\begin{tabular}{||l||}
\hline\hline
$S_{B}=\ln W$. \\ \hline\hline
\end{tabular}
\label{bp}
\end{equation}
The Boltzmann's Principle constitutes the starting point of the
microcanonical thermostatistics theory of D. H. E. Gross (see the refs.\cite
{gro1,gro2,gro3}). From here is concluded that the pseudoextensive systems
are the natural frame for the application of this formulation.

\section{The Legendre formalism}

In this section it will be performed the analysis of the equivalence of the
microcanonical ensemble with some generalized canonical ensemble in analogy
with the work done by D. H. E. Gross in deriving his microcanonical
thermostatistics\cite{gro1,gro2,gro3}. Following Boltzmann, all the
statistical information about the general properties of the macroscopic
state of any Hamiltonian system is contained in the accessible
configurational space volume of the microcanonical ensemble, $W\left(
I,N,a\right) $:

\begin{equation}
W\left( I,N,a\right) =\Omega \left( I,N,a\right) \delta I_{o}=\delta
I_{o}\int \delta \left[ I-I_{N}\left( X;a\right) \right] dX
\end{equation}
where $\delta I_{o}$ is a {\it suitable} constant volume element, which
makes $W$ dimensionless. For the pseudoextensive systems the ordering
information is contained by the function $\gamma \left( I,N,a\right) $:

\begin{equation}
\gamma \left( I,N,a\right) \equiv \frac{1}{N^{\kappa }}\ln W\left(
I,N,a\right) \text{,}
\end{equation}
which is scaling invariant when ${\cal R}_{I}\in {\cal M}_{c}$ and the ThL
is reached.

In this case, the corresponding information entropy is the very well known
Shannon-Boltzmann-Gibbs' extensive entropy (SBGE):

\begin{equation}
S_{SBG}=-%
\mathrel{\mathop{\sum }\limits_{k}}%
p_{k}\ln p_{k}\text{.}
\end{equation}

In the thermodynamic equilibrium the SBGE leads to the ordinary canonical
ensemble where its PDF are known as the Boltzmann-Gibbs' Distributions: 
\begin{equation}
\omega \left( X;\beta ,N,a\right) =\frac{1}{Z\left( \beta ,N,a\right) }\exp %
\left[ -\beta \cdot I_{N}\left( X;a\right) \right] \text{,}
\end{equation}
where $\beta $ are the canonical parameters and $Z\left( \beta ,a,N\right) $
is the partition function. The states density of the microcanonical ensemble
and the partition function of the canonical ensemble are related by means of
the {\it Laplace Transformation:}

\begin{equation}
Z\left( \beta ,N,a\right) =\int \exp \left[ -\beta \cdot I\right] W\left(
I,N,a\right) \frac{dI}{\delta I_{o}}\text{,}
\end{equation}
which establishes the connection between the Boltzmann's entropy with the
fundamental thermodynamics potential of the canonical distribution, {\it the
Planck Potential}:

\begin{equation}
P\left( \beta ,N,a\right) =-\ln Z\left( \beta ,N,a\right)
\end{equation}

The Planck Potential is adopted since it does not introduce any preference
with an specific integral of movement of the microcanonical distribution, in
order to be consequent with the $SL({\bf R}^{n})$ invariance. The last
integral can be rewritten as:

\begin{equation}
\exp \left[ -P\left( \beta ,N,a\right) \right] =\int \exp \left[ -\beta
\cdot I+S_{B}\left( I,N,a\right) \right] \frac{dI}{\delta I_{o}}\text{.}
\end{equation}

Immediately, it is recognized in the exponential argument the {\it %
Legendre's Transformation} between the thermodynamic potentials. If the
integrals of movements and the Boltzmann's entropy has the same scaling
transformation in the ThL, that is:

\begin{equation}
\chi =\kappa \text{,}  \label{c2}
\end{equation}
this integral will have a very sharp peak around the maximum value,{\it \ \
if this maximum exists} , and therefore, the main contribution to this
integral will come from it. Thus, for the equivalence between the
microcanonical and the canonical ensembles only are possible those
representations $\Re _{I}$ that satisfy the condition given by Eq.(\ref{c2}).

When the equivalence is held, the Planck potential could be approached by
means of the Legendre's formalism:

\begin{equation}
P\left( \beta ,N,a\right) \simeq 
\mathrel{\mathop{\max }\limits_{I}}%
\left[ \beta \cdot I-S_{B}\left( I,N,a\right) \right] \text{,}
\end{equation}
where $\beta $ is obtained by: 
\begin{equation}
\beta _{\mu }=\left. \frac{\partial }{\partial I^{\mu }}S_{B}\right|
_{I=I_{M}}\text{.}
\end{equation}

For the stability of the maximum is necessary that all the eingenvalues of
the {\it curvature tensor}:

\[
\left( K_{B}\right) _{\mu \nu }=\left. \frac{\partial }{\partial I^{\mu }}%
\frac{\partial }{\partial I^{\nu }}S_{B}\right| _{I=I_{M}}, 
\]
be negatives, that is, the Boltzmann's entropy must be locally concave
around the maximum. In this case, in the canonical ensemble there will be
small fluctuations of the integrals of movement around its mean values with
a standard desviation:

\begin{equation}
\frac{\sqrt{\overline{\left( I^{\mu }-I_{M}^{\mu }\right) \left( I^{\nu
}-I_{M}^{\nu }\right) }}}{I_{M}^{\mu }I_{M}^{\nu }}\sim \frac{1}{\alpha ^{%
\frac{\kappa }{2}}}\text{.}
\end{equation}

However, if the concavity condition for the entropy is not hold, there will
be a catastrophe in the Laplace transformation and the canonical ensemble
will not be able to describe the system. In the general sense, this
situation is a sign of the occurrence of generalized phase transitions.

The scaling behavior of the system impose one restriction to the macroscopic
observables of the systems, the generalized Duhem-Gibbs' relation:

\begin{equation}
\kappa \left( \beta \cdot I^{\ast }-S_{B}\right) +\mu N+\pi _{a}\beta \cdot
f_{a}a=0\text{,}
\end{equation}
where:

\begin{equation}
\mu =\frac{\partial }{\partial N}S_{B}\text{, \ }\beta \cdot f_{a}=\frac{%
\partial }{\partial a}S_{a}\text{ with }f_{a}=\left\langle -\frac{\partial }{%
\partial a}I_{N}^{\ast }\left( X;a\right) \right\rangle \text{.\ }
\end{equation}

Although the pseudoextensive systems are nonextensive in the usual sense,
their study could be carried out with the same formalism used for the
ordinary extensive systems with appropriate selection of the representation
of the movement integrals space. The Eq.(\ref{c2}) leads to condition of the 
{\it homogeneous scaling of the movement integrals and the entropy} in order
to satisfy the requirement of the scaling invariance of the canonical
parameters, the validity of the zero principle of the thermodynamics. This
is the cause of the spontaneous symmetry breaking. Although in the
microcanonical ensemble all the representations of the movement integrals
are equivalent, in the generalized canonical ensemble only are possible
those representations with an {\it homogeneous nondegenerated scaling}. To
understand the term {\it nondegenerate }\ let us show the following example.
Let be $A$ and $B$ two integrals of movement with different scaling behavior:

\begin{equation}
A\sim \alpha ^{a}\text{,\ }B\sim \alpha ^{b}\text{ with }a>b\text{.}
\end{equation}
In another representation, these integrals could be equivalently represented
in the microcanonical ensemble by:

\begin{equation}
I^{\pm }=A\pm B\text{,}
\end{equation}
but, in this case, their scaling behavior will be given by: 
\begin{equation}
I^{\pm }\approx A\sim \alpha ^{a}\text{.}
\end{equation}

$I^{\pm }$ have a homogeneous scaling in the ThL, but they are not
independent. The canonical parameters derived from $I^{\pm }$, $\beta ^{\pm
}=\frac{\partial S_{B}}{\partial I^{\pm }}$, will be identically in the ThL.
This leads to the {\it trivial vanishing }of the {\it curvature tensor
determinant }:

$%
\mathrel{\mathop{\lim }\limits_{\alpha \rightarrow \infty }}%
\alpha ^{2a}\det \left| \left( K_{B}\right) _{\mu \nu }\right| =$%
\begin{equation}
\mathrel{\mathop{\lim }\limits_{\alpha \rightarrow \infty }}%
\alpha ^{2a}\det \left( 
\begin{tabular}{ll}
$\partial _{+}\partial _{+}S_{B}$ & $\partial _{-}\partial _{+}S_{B}$ \\ 
$\partial _{+}\partial _{-}S_{B}$ & $\partial _{-}\partial _{-}S_{B}$%
\end{tabular}
\right) \equiv 0\text{.}
\end{equation}

This is an example of an\ homogeneous degenerate scaling representation.
These representations are inadmissible for the generalized canonical
ensemble. The above considerations give a criterium of\ an homogeneous
nondegenerated scaling representation, the non trivial vanishing of the
curvature tensor determinant:

\begin{equation}
\mathrel{\mathop{\lim }\limits_{\alpha \rightarrow \infty }}%
\det \left( \alpha ^{\kappa }\partial _{\mu }\partial _{\nu }S_{q}\right)
\neq 0
\end{equation}

A final remark: In general, each a representation ${\cal R}_{I}$ belonging
to ${\cal M}_{c}$ will posses an specific region of $\Im _{N}$ in which will
be equivalent the microcanonical and the canonical ensemble. This fact
reflects the inconsistency of the above analysis because the description of
the system should not depend on the representation. It could be corrected
introducing a local reparametrization invariant formalism, maybe,
introducing the {\em covariant derivative}.

\section{Conclusions}

We have showed that the pseudoextensive systems could be described by the
ordinary Boltzmann-Gibbs' Statistic with an appropriate selection of the
representation of the movement integrals space of the microcanonical
ensemble, $\Im _{N}$. The key for this conclusion is the assuming of scaling
postulates: The equivalence of the microcanonical description with a
generalized canonical one during the realization of the Thermodynamic Limit
by means of the self-similarity scaling properties of the fundamental
physical observables. It is shown that the pseudoextensive systems are the
natural frame for the application of the microcanonical thermostatistics
theory of D. H. E. Gross.

The general lines of the present analysis is analogue to the method used by
D. H. E. Gross in deriving the thermodynamic formalism of his formulation.
However our derivation clarifies some incongruences present in this theory.
The Gross' theory does not take into account the reparametrization
invariance of the microcanonical ensemble neither the scaling properties of
the nonextensive systems in general way. In its present status, this theory
could only be valid for those systems that in the ThL converge to an
ordinary extensive systems, since its formalism is based on the
consideration of the additive representations of the movement integrals (see
for example in refs.\cite{gro4,gro5}).

\onecolumn

\end{document}